\documentclass[aps,prb,twocolumn,showpacs,superscriptaddress,floatfix]{revtex4-1}
\usepackage{dcolumn}
\usepackage{multirow}
\usepackage{amsmath}
\usepackage{graphicx,epsfig,psfrag}
\usepackage{amssymb}
\usepackage{natbib}
\usepackage[colorlinks,linkcolor={magenta},citecolor={magenta},urlcolor={magenta},breaklinks=true]{hyperref}
\usepackage{bookmark}

\begin{document}

\title{Full density matrix numerical renormalization group calculation 
of impurity susceptibility and specific heat of the Anderson impurity model}
 
\author{L. Merker}
\affiliation
{Peter Gr\"{u}nberg Institut and Institute for Advanced Simulation, 
Research Centre J\"ulich, 52425 J\"ulich, Germany}
\author{A. Weichselbaum}
\affiliation{Physics Department, Arnold Sommerfeld Center for Theoretical Physics, and Center for NanoScience,
Ludwig-Maximilians-Universit\"{a}t, Theresienstra{\ss}e 37, D-80333 M\"{u}nchen, Germany}
\author{T. A. Costi}
\affiliation
{Peter Gr\"{u}nberg Institut and Institute for Advanced Simulation, 
Research Centre J\"ulich, 52425 J\"ulich, Germany}

\begin{abstract}
Recent developments in the numerical renormalization group (NRG) allow the
construction of the full density matrix (FDM) of quantum impurity
models (see A. Weichselbaum and J. von Delft in Ref.~\onlinecite{Weichselbaum2007}) 
by using the completeness of the eliminated states introduced by 
F.\, B.\, Anders and A.\, Schiller in Ref.~\onlinecite{Anders2005}. 
While these developments prove particularly useful in the
calculation of transient response and finite temperature Green's functions of
quantum impurity models, they may also be used to calculate thermodynamic properties.
In this paper, we assess the FDM approach to thermodynamic
properties by applying it to the Anderson impurity model.
We compare the results for the susceptibility and specific heat 
to both the conventional approach within NRG and to exact Bethe ansatz results. 
We also point out a subtlety in the calculation of the susceptibility (in a uniform
field) within the FDM approach. Finally, we show numerically that for the Anderson model,
the susceptibilities in response to a local and a uniform magnetic field coincide in
the wide-band limit, in accordance with the Clogston-Anderson compensation theorem.
\end{abstract}

\pacs{75.20.Hr, 71.27.+a, 72.15.Qm}


\date{\today}

\maketitle

\section{Introduction} 
\label{sec:introduction}

The numerical renormalization group method, 
\cite{Wilson1975,KWW1980a,KWW1980b,Bulla2008} has proven very successful 
for the study of quantum impurity models.\cite{Hewson1997}. Initially 
developed to describe, in a controlled non-perturbative fashion, 
the full crossover from weak to strong coupling behavior in the Kondo 
problem \cite{Wilson1975} and the temperature dependence of the impurity thermodynamics
,\cite{Wilson1975,KWW1980a,KWW1980b,Oliveira1981} it has subsequently been 
extended to dynamic \cite{Frota1986,Sakai1989,Costi1992} 
and transport properties \cite{Costi1994} of quantum impurity models. 
Recently, a number of refinements to the calculation of dynamic properties have
been made, including the use of the correlation self-energy in evaluating 
Green functions \cite{Bulla1998}, the introduction of the reduced density 
matrix \cite{Hofstetter2000}, and the introduction of a complete basis 
set using the eliminated states in each NRG iteration 
\cite{Anders2005}. The latter, in combination with the reduced density matrix, has been 
used to evaluate the multiple shell summations arising in the time dependent transient 
response in quantum impurity problems \cite{Costi1997,Anders2005}  
and offers the possibility to investigate truly non-equilibrium
steady state transport within the NRG method \cite{Anders2008}. In addition, the complete
basis set offers an elegant way to calculate finite temperature Green functions 
which satisfy the fermionic sum rules exactly.\cite{Weichselbaum2007,Peters2006,Toth2008} For recent
applications of this technique to transport properties, see 
Refs.~\onlinecite{Costi2009,Costi2010}.

In this paper we benchmark the full density matrix (FDM) approach to thermodynamic properties, 
by applying it to the prototype model of strong correlations, the Anderson impurity model \cite{Anderson1961}.
This model has been solved exactly using the Bethe ansatz. 
\cite{Kawakami1981,Kawakami1982a,Okiji1983,Wiegmann1983,Tsvelick1983,Filyov1982,Tsvelick1982}
A numerical solution of the resulting thermodynamic Bethe ansatz (TBA) equations therefore allows
one to compare the FDM results for quantities such as the specific heat and the
susceptibility with essentially exact calculations from the Bethe ansatz. In addition, 
we shall also compare the FDM results for specific heats and susceptibilities with those of
the conventional approach \cite{Campo2005} (see Sec.~\ref{sec:conventional} for a more precise
definition of what we term ``conventional'').

The paper is organized as follows. In Section~\ref{sec:conventional} we
specify the Anderson impurity model, and outline the conventional approach to thermodynamics within the
NRG method \cite{Campo2005}. The FDM approach to thermodynamics is described in Section~\ref{sec:fdm}.
Section~\ref{sec:results} contains our results. The impurity contribution to the specific heat, $C_{\rm imp}$, 
calculated within the FDM approach, is compared with Bethe ansatz calculations 
\cite{Kawakami1981,Kawakami1982a,Okiji1983,Wiegmann1983,
Tsvelick1983,Filyov1982,Tsvelick1982} and to calculations using the conventional approach
in Sec.~\ref{subsec:specific heat}. The impurity contribution to the susceptibility, $\chi_{\rm imp}$, 
with the magnetic field acting on both impurity and conduction electron states, calculated within
FDM is compared with corresponding results from Bethe ansatz and the conventional approach within NRG
in  Sec.~\ref{subsec:susceptibility}). Results for the Wilson ratio as a function of the local Coulomb 
repulsion and the local level position are also given in  Sec.~\ref{subsec:susceptibility}).
In Sec.~\ref{subsec:local susceptibility} we consider also the local
susceptibility of the Anderson model, $\chi_{\rm loc}$, with a magnetic field acting only on the impurity,
and show by comparison with Bethe ansatz results for $\chi_{\rm imp}$, that $\chi_{\rm loc}=\chi_{\rm imp}$ for
both the symmetric and asymmetric Anderson model in the wide-band limit.
In addition, we also compare the FDM and conventional approaches for another local quantity, the
double occupancy, in Sec.~\ref{subsec:double occupancy}. Section~\ref{sec:summary} contains our summary.
Details of the numerical solution of the thermodynamic Bethe ansatz equations 
may be found in Ref.~\onlinecite{Merker2012}. 
\section{Model, method and conventional approach to thermodynamics}
\label{sec:conventional}
We consider the Anderson impurity model \cite{Anderson1961} in a magnetic field $B$, 
described by the Hamiltonian  $$H  = H_{\rm imp} + H_{\rm 0} + H_{\rm int} + H_{B}.$$
The first term, 
$H_{\rm imp}  = \sum_{\sigma}\varepsilon_{d}d_{\sigma}^{\dagger}d_{\sigma} 
+ Un_{d\uparrow}n_{d\downarrow}$, describes the impurity with local level
energy $\varepsilon_{d}$ and onsite Coulomb repulsion $U$, the second term,
$H_{\rm 0} =  \sum_{k\sigma}\epsilon_{k}c_{k\sigma}^{\dagger}c_{k\sigma}$, 
is the kinetic energy of non-interacting conduction electrons with 
dispersion $\varepsilon_{k}$, 
the third term, $H_{\rm int}  =  V\sum_{k\sigma}(c_{k\sigma}^{\dagger}d_{\sigma}
+d^{\dagger}_{\sigma}c_{k\sigma})$, is the hybridization between the local level
and the conduction electron states, with $V$ being the hybridization matrix element, and,
the last term $H_{B}=-g\mu_{\rm B}B\,S_{z,tot}$ 
where $S_{z,tot}$ is the $z$-component of the total spin (i.e. impurity plus 
conduction electron spin), is the uniform magnetic field acting on impurity and
conduction electrons. $g$ is the electron g-factor, and $\mu_{\rm B}$ is the 
Bohr magneton. We choose units such that $g=\mu_{\rm B}=1$ and assume a constant
conduction electron density of states per spin $N(\epsilon)=1/2D$, where $D=1$ is
the half-bandwidth. The hybridization strength is denoted by 
$\Delta_{0}=\pi V^{2}N(0)$ and equals the half-width of the non-interacting resonant
level.

The NRG procedure\cite{KWW1980a,KWW1980b,Bulla2008}  
consists of iteratively diagonalizing a discrete form of the
above Hamiltonian $H$. It starts out by replacing the quasi-continuum of 
conduction electron energies $-D\le \varepsilon_{k}\le D$ by 
logarithmically discretized ones about the Fermi level $\varepsilon_{F}=0$, i.e. 
$\epsilon_{n} = \pm D\Lambda^{-n -(1-z)}, n=1,\dots$, where
$\Lambda>1$ is a rescaling factor. Averaging physical quantities over
several realizations of the logarithmic grid, defined by the
parameter $z\in (0,1]$, eliminates artificial discretization induced 
oscillations at $\Lambda\gg 1$. \cite{Oliveira1994,Campo2005,Zitko2009b}
Rotating the discrete conduction states into a Wannier basis
$f_{n\sigma},n=0,1,2,\dots$ at the impurity site, one arrives at the
form 
$H=\lim_{m\rightarrow\infty} H_{m}$, where the truncated Hamiltonians
$H_{m}$ are defined by
$H_{m} = H_{\rm imp} + H_{\rm hyb} + \sum_{n=0\sigma}^{m}\tilde{\epsilon}_{n}(z)
f_{n\sigma}^{\dagger}f_{n\sigma} 
+ \sum_{n=0\sigma}^{m-1}t_{n}(z)(f_{n\sigma}^{\dagger}f_{n+1\sigma}+
f_{n+1\sigma}^{\dagger}f_{n\sigma}),$
with $H_{\rm imp}$ as defined previously and $H_{\rm
  hyb}=V\sum_{\sigma}(f_{0\sigma}^{\dagger}d_{\sigma}+d_{\sigma}^{\dagger}f_{0\sigma})$.
The onsite energies $\tilde{\epsilon}_{n}(z)$ and hoppings
$t_{n}(z)$ reflect the energy dependence of the hybridization
function and density of states.\cite{KWW1980a,KWW1980b,Bulla2008} The sequence of
truncated Hamiltonians $H_{m}$ is then iteratively diagonalized by using the recursion relation 
$H_{m+1}=H_{m}+\sum_{\sigma}\tilde{\epsilon}_{m}(z)f_{m\sigma}^{\dagger}f_{m\sigma} +
\sum_{\sigma}t_{m}(z)(f_{m\sigma}^{\dagger}f_{m+1\sigma}+f_{m+1\sigma}^{\dagger}f_{m\sigma})$.
The resulting eigenstates $|p;m\rangle$ and 
eigenvalues $E_{p}^{m}$, obtained  on a decreasing set of energy scales 
$\omega_{m}(z)\sim t_{m}(z), m=0,1,\dots$, are then used to obtain
physical properties, such as Green's functions or thermodynamic properties.
Unless otherwise stated, we use conservation of total electron 
number $N_{e}$, total spin $S$, and total $z$-component of spin $S_{z}$ 
in the iterative diagonalization of $H$ at $B=0$, so the
eigenstate $|p;m\rangle$ is an abbreviation for the eigenstate 
$|N_{e}SS_{z}p;m\rangle$ of $H_{m}$, with energy $E_{N_{e}Sp}^{m}$
(abbreviated as $E_{p}^{m}$) where the index $p=1,\dots$ distinguishes states with
the same conserved quantum numbers. As long as $m\le m_{0}-1$, 
{
where typically $m_{0}=4-6$
}
, all states are retained. For $m\ge m_{0}$, only the lowest energy states
are used to set up the Hamiltonian $H_{m+1}$. These may be a fixed
number $N_{\rm keep}$ of the lowest energy states, or one may specify a predefined
$m_{0}$, and retain only those states with rescaled energies 
$(E_{p}^{m}-E^{m}_{GS})/t_{m}(z) < e_{c}(\Lambda)$, where $E_{GS}^{m}$ 
is the (absolute) groundstate energy at iteration $m$ and 
$e_{c}(\Lambda)$ is $\Lambda$-dependent cut-off energy. \cite{Costa1997,Oliveira1994,Weichselbaum2011}
{For most results in this paper, we used $m_{0}=4,5$,
  respectively for $H$, $H_{0}$, and $e_{c}(\Lambda)=15\sqrt{\Lambda}\approx 47$ for $\Lambda=10$
and found excellent agreement with exact continuum results from Bethe ansatz (after appropriate
$z$-averaging, see below). Calculations at smaller $\Lambda=4$, using
$m_{0}=5,6$ for  $H$, $H_{0}$, respectively, and $e_{c}(\Lambda)=40$ were also
carried out for the local susceptibility in Sec.~\ref{subsec:local susceptibility}. These showed equally 
good agreement with corresponding continuum Bethe ansatz results,
indicating that the $m_{0}$ and $\Lambda$-dependence of our results (after $z$-averaging) is negligible.
In our notation, the number of retained states at iteration $m$ (before truncation) grows as $4^{m+1}$,
so the value of $m_{0}$ cannot be increased much beyond $5$, in practice, due to the exponential increase
in storage and computer time. As our calculations show, this is also not necessary, since agreement with
exact Bethe ansatz calculations is achieved already for $m\ge m_{0}=4,5$.
}

The impurity contribution to the specific heat is defined by $C_{\rm imp}(T)=C(T)-C_{0}(T)$, where
$C(T)$ and $C_{0}(T)$ are the specific heats of $H$ and $H_{0}$, respectively. Similarly, 
the impurity contribution to the zero field susceptibility is given by
$\chi_{\rm imp}(T)=\chi(T)-\chi_{0}(T)$, where $\chi(T)$ and $\chi_{0}(T)$ are the
susceptibilities of $H$ and $H_{0}$, respectively. Denoting by 
{$Z(T,B)$ 
and $Z_{\rm 0}(T,B)$ the 
partition functions
}
of $H$ and $H_{0}$, and $\Omega(T,B)=-k_{\rm B}T \ln Z(T,B)$ 
and $\Omega_{0}(T,B)=-k_{\rm B}T\ln Z_{\rm 0}(T,B)$ the corresponding thermodynamic potentials,
we have, 
\begin{eqnarray}
C(T) &=&-T\frac{\partial^{2}\Omega(T)}{\partial\;T^{2}}
=k_{\rm B}\beta^{2}\langle (H-\langle\; H\rangle)^{2} \rangle,\label{spec-total}\\
C_{0}(T) &=&-T\frac{\partial^{2}\Omega_{0}(T)}{\partial\;T^{2}}
=k_{\rm B}\beta^{2}\langle (H_{0}-\langle\; H_{0}\rangle)^{2} \rangle_{0},\label{spec-host}\\
\chi(T) &=&-\frac{\partial^{2}\Omega(T,B)}{\partial\;B^{2}}|_{B=0}
=\beta(g\mu_{\rm B})^{2}\langle S_{z,tot}^{2} \rangle,\label{chi-total}\\
\chi_{0}(T) &=&-\frac{\partial^{2}\Omega_{0}(T,B)}{\partial\;B^{2}}|_{B=0}
=\beta (g\mu_{\rm B})^{2}\langle {S_{z,tot}^{0}}^{2} \rangle_{0}\label{chi-host},
\end{eqnarray}
where $S_{z,tot}^{0}$ is the $z$-component of total spin for $H_{0}$.
 
We follow the approach of Ref.~\onlinecite{Campo2005}, 
which we term the ``conventional'' approach, to calculate the thermodynamic
averages appearing in (\ref{spec-total}-\ref{chi-host}) at large $\Lambda\gg 1$
(thermodynamic calculations at smaller values of $\Lambda\lesssim 3$ are also
possible,\cite{KWW1980a,Costi1994} however, truncation errors increase with
decreasing $\Lambda$): for any temperature $T$, we choose the smallest $m$ such that
$k_{B}T > t_{m}(z)$ and we use the eigenvalues of $H_{m}$ to evaluate 
the partition function
$Z_{m}(T)=\sum_{p}e^{-E_{p}^{m}/k_{\rm B}T}$ and the 
expectation values appearing in  (\ref{spec-total}-\ref{chi-host}).
Calculations for several $z=(2i-1)/2n_{z}, i=1,\dots,n_{z}$ with typically
$n_{z}=2,4$ or $8$ are carried out and averaged in order to eliminate discretization
induced oscillations at large $\Lambda\gg 1$. A dense grid of temperatures defined
on a logarithmic scale from $10^{-4}T_{\rm K}$ to $2D$ was used throughout, where $T_{K}$
is the Kondo scale for the symmetric Anderson model for given $U$ 
[see Eq.~(\ref{eq:symmetric-TK})].  
{
An advantage of the FDM approach,
which we describe next, is that such a dense grid of temperatures can
be used without the requirement to choose a best shell for a given
$T$ and $z$. This is possible within the FDM approach, because the 
partition function of the latter contains all excitations from all shells.
}

\section{Thermodynamics within the FDM approach}
\label{sec:fdm}

An alternative approach to thermodynamics is offered by making use of the eliminated states \cite{Anders2005}
from each NRG iteration. These consist of 
the set of states $|lem\rangle=|lm\rangle|e\rangle$ obtained from the eliminated 
eigenstates, $|lm\rangle$, of $H_{m}$, and the degrees of freedom,
denoted collectively by $e$, of the sites $i=m+1,\dots,N$,
where $N$ is the longest chain diagonalized. The set of states $|lem\rangle$ for $m=m_{0},\dots,N$
form a complete set, with completeness being expressed by \cite{Anders2005}
\begin{eqnarray}
1&=&\sum_{m'=m_{0}}^{N}\sum_{le}|lem'\rangle\langle lem'|
\label{completeness}
\end{eqnarray}
where $m_{0}-1$ is the last iteration for which all states are retained. 
Weichselbaum and von Delft \cite{Weichselbaum2007} introduced the full density 
matrix (FDM) of the system made up of 
the complete set of eliminated states from all iterations $m=m_{0}+1,\dots,N$. 
Specifically, the FDM is defined by
\begin{eqnarray}
\rho = \sum_{m=m_{0}}^{N}\sum_{le}|lem\rangle \frac{e^{-\beta E_{l}^{m}}}{Z(T)} \langle lem|
\label{fdm}
\end{eqnarray}
where $Z(T)$ is the partition function made up from the complete spectrum, i.e. it contains
all eliminated states from all $H_{m}, m=m_{0},\dots,N$ [where all states of the last
iteration $m=N$ are included as eliminated states, so that Eq.~(\ref{completeness}) holds]. 
Similarly, one may define the full density matrix, $\rho_{0}$, for the host system $H_{0}$, by 
\begin{eqnarray}
\rho_{0} = \sum_{m=m_{0}'}^{N}\sum_{le}|lem\rangle \frac{e^{-\beta E_{l,0}^{m}}}{Z_{0}(T)} \langle lem|,
\label{fdm0}
\end{eqnarray}
where  $Z_{0}(T)$ is the full partition function of $H_{0}$. Note that $m_{0}'$ may differ from $m_{0}$, 
as the impurity site is missing from $H_{0}$. In order to evaluate the
thermodynamic average of an operator $\hat{O}$ with respect to the FDM of equation~(\ref{fdm}), we
follow Weichselbaum and von Delft~\cite{Weichselbaum2007} and introduce the normalized density
matrix for the m'th shell in the Hilbert space of $H_N$:
\begin{equation}
\tilde{\rho}_{m} = \sum_{le}|lem\rangle \frac{e^{-\beta E_{l}^{m}}}{\tilde{Z}_{m}} \langle lem|.
\end{equation}
Normalization, ${\rm Tr}[\tilde{\rho}_{m}]=1$, implies that
\begin{equation}
1 = \sum_{l}\frac{e^{-\beta E_{l}^{m}}}{\tilde{Z}_{m}}d^{N-m}=d^{N-m}\frac{Z_{m}}{\tilde{Z}_{m}},
\end{equation}
where $Z_{m}=\sum_{l}e^{-\beta E_{l}^{m}}$ and 
{$\tilde{Z}_{m}=d^{N-m}Z_{m}$ with the factor $d^{N-m}$ resulting
from the trace over the $N-m$ environment degrees of freedom $e\equiv (e_{m+1},e_{m+2},\dots,e_{N})$. 
For the single channel Anderson model, considered here, $d=4$, since each $e_{i}$ assumes
four possible values (empty, singly occupied up/down and doubly occupied states).
}
Then the FDM can be written as a sum of weighted density matrices for shells $m=m_{0},\dots,N$
\begin{eqnarray}
\rho &=& \sum_{m=m_{0}}^{N}w_{m}\tilde{\rho}_{m},\\
w_{m} &=& d^{N-m}\frac{Z_{m}}{Z}\label{weights},
\end{eqnarray}
where $\sum_{m=m_{0}}^{N}w_{m} = 1$ and the calculation of the weights $w_{m}$ is 
outlined in Ref.~\onlinecite{Costi2010}.

Substituting $\rho=\sum_{m'}w_{m'}\tilde{\rho}_{m'}$ into the expression for the
thermodynamic average $\langle \hat{O}\rangle$ and making use of the decomposition of unity 
Eq.~(\ref{completeness}), we have
\begin{eqnarray}
\langle \hat{O}\rangle_{\rho} &=& {\rm Tr}\left[ \rho \hat{O}\right]\nonumber\\
&=& \sum_{l'e'm'}\langle l'e'm'|\hat{O} \sum_{lem} w_{m}|lem\rangle\frac{e^{-\beta E_{l}^{m}}}{\tilde{Z}_{m}}
\langle lem|l'e'm'\rangle\nonumber\\
&=&\sum_{lem}O_{ll}^{m}w_{m}\frac{e^{-\beta E_{l}^{m}}}{\tilde{Z}_{m}}\label{fdm-intermediate}\\
&=&\sum_{lm}d^{N-m}w_{m}O_{ll}^{m}\frac{e^{-\beta E_{l}^{m}}}{d^{N-m}{Z}_{m}}\nonumber\\
&=&\sum_{m=m_{0},l}^{N}w_{m}O_{ll}^{m}\frac{e^{-\beta E_{l}^{m}}}{{Z}_{m}},\label{fdm-average}
\end{eqnarray}
where orthonormality $\langle lem|l'e'm'\rangle=\delta_{ll'}\delta_{ee'}\delta_{mm'}$, and
the trace over the $N-m$ environment degrees of freedom $\sum_{lem}\dots = \sum_{lm}d^{N-m}\dots$ has
been used and $O_{ll}^{m}=\langle lm|\hat{O}|lm\rangle$. A similar expression applies for expectation
values $\langle \hat{O}\rangle_{\rho_{0}}$ with respect to the host system $H_{0}$.
For each temperature $T$ and shell $m$, 
we require $w_{m}(T)$ and the factor $B_{l}^{m}(T)=e^{-\beta E_{l}^{m}}/Z_{m}$ 
where $Z_{m}=\sum_{l}e^{-\beta E_{l}^{m}}$. Numerical problems due to large exponentials
are avoided by calculating $B_{l}^{m}(T)=e^{-\beta (E_{l}^{m}-E_{0}^{m})}/Z_{m}'$ where 
$Z_{m}'=e^{\beta E_{0}^{m}}Z_{m}$ and $E_{0}^{m}$ is the lowest energy for shell $m$.

The calculation of the full partition function $Z(T)$ , like the weights $w_{m}(T)$, requires care
in order to avoid large exponentials (see Ref.~\onlinecite{Costi2010}). Note also, that
the energies $E_{l}^{m}$ in the above expressions denote absolute energies of $H_{m}$. In practice, in NRG
calculations one defines rescaled Hamiltonians $\bar{H}_{m}$ in place of $H_{m}$, with rescaled
energies $\bar{E}_{l}^{m}$ shifted so that the groundstate energy of $\bar{H}_{m}$ is zero. In the
FDM approach, information from different shells is combined. This requires that energies
from different shells be measured relative to a common groundstate energy, which is usually 
taken to be the absolute groundstate energy of the longest chain diagonalized. 
Hence, it is important to keep track of rescaled groundstate energies of the $\bar{H}_{m}$ 
so that the $\bar{E}_{l}^{m}$ can be related to the absolute energies $E_{l}^{m}$ used in the FDM
expressions for thermodynamic averages (this relation is specific to precisely how the 
sequence $\bar{H}_{m}$, $m=1,2,\dots$ is defined, so we do not specify it here).

The specific heat $C_{\rm imp}(T)=C(T)-C_{0}(T)$ is obtained from separate calculations
for $H$ and $H_{0}$. For $H$, we first calculate $E=\langle H\rangle$ using (\ref{fdm-average}):
\begin{eqnarray}
\langle H\rangle_{\rho} &=&\sum_{m=m_{0},l}^{N}w_{m}E_{l}^{m}B_{l}^{m},\label{h-average}
\end{eqnarray}
and then substituting this into equation~(\ref{spec-total}) to obtain
\begin{eqnarray}
C(T)&=&k_{\rm B}\beta^{2}\langle (H-\langle\; H\rangle)^{2} \rangle\nonumber\\
&=&k_{\rm B}\beta^{2}\sum_{m=m_{0},l}^{N}w_{m}(E_{l}^{m}-E)^{2}B_{l}^{m},
\label{spec-total-fdm}
\end{eqnarray}
with a similar calculation for $H_{0}$ to obtain $C_{0}(T)$. The specific heats are
then $z$-averaged and subtracted to yield $C_{\rm imp}(T)=C(T)-C_{0}(T)$. Alternatively,
the specific heat $C(T)$ may be obtained from the $z$-averaged entropy $S(T)$ via 
$C(T)=-T\partial S/\partial T$, where $S$ is calculated from $E$ and $Z$ using 
$S=-\partial \Omega/\partial T = k_{\rm B}\ln Z + E/T$ (with similar expressions for
$C_{0}(T)$ and $S_{0}(T)$). We note that in cases where explicit numerical derivatives 
of the thermodynamic potential with respect to magnetic field or temperature are 
required, the NRG supplies a sufficiently smooth $\Omega(T,B)$ for this to be possible 
(see Ref.~\onlinecite{Costi1994} for an early application). 
We show this within the FDM approach for the case of the local magnetic susceptibility
in Sec.~\ref{subsec:local susceptibility}, a quantity that requires
a numerical second derivative of $\Omega(T,B)$ with respect to $B$.

The susceptibility from (\ref{chi-total}-\ref{chi-host}) requires more care, since a uniform
field acts also on the environment degrees of freedom, implying that we require the expectation
value $\langle (S_{z}+S_{z,e})^{2}\rangle$ in evaluating $k_{\rm B}T\chi(T,B=0)/(g\mu_{B})^{2}$
(and similarly for $k_{\rm B}T\chi_{0}(T,B=0)/(g\mu_{B})^{2}$), where $S_{z}$ refers to total
$z$-component of spin for the system $H_{m}$ and $S_{z,e}$, the total $z$-component of
the $N-m$ environment states $e$. Now 
$\langle (S_{z}+S_{z,e})^{2}\rangle = \langle (S_{z}^{2}+S_{z,e}^{2})\rangle$ since the trace
over $S_{z}$ (or $S_{z,e}$) of the cross-term $2S_{z}S_{z,e}$ will vanish. Hence, the susceptibility
will have an additional contribution $\chi_{\rm E}=\beta (g\mu_{\rm B})^{2}\langle S_{z,e}^{2}\rangle$ due the
environment degrees of freedom in addition to the usual term 
$\chi_{\rm S}=\beta (g\mu_{\rm B})^{2}\langle S_{z}^{2}\rangle$ for the system $H_{m}$. 
Evaluating the latter via (\ref{fdm-average}), indicating explicitly the conserved quantum 
number $S_{z}$ in the trace with all other conserved quantum numbers indicated by $l$, results in
\begin{eqnarray}
\frac{k_{\rm B}T\chi_{\rm S}}{(g\mu_{B})^{2}}
&=&\langle S_{z}^{2} \rangle =\sum_{m=m_{0},S_{z},l}^{N}
w_{m}S_{z}^{2}B_{l}^{m}\nonumber\\
&=&\sum_{m=m_{0},l}^{N}f_{1}(S)w_{m}B_{l}^{m}
\end{eqnarray}
where $\sum_{S_{z}}S_{z}^{2}=f_{1}(S)=(2S+1)((2S+1)^{2}-1)/12$ has been used.
For the term $\chi_{\rm E}$, we have from (\ref{fdm-intermediate})
\begin{eqnarray}
\frac{k_{\rm B}T\chi_{\rm E}}{(g\mu_{B})^{2}}
&\equiv &\langle S_{z,e}^{2}\rangle
=\sum_{m=m_{0},S_{z},l,e}^{N}w_{m}S_{z,e}^{2}\frac{e^{-\beta E_{l}^{m}}}{\tilde{Z}_{m}}.
\end{eqnarray}
Since $\tilde{Z}_{m}=d^{N-m}Z_{m}$ and denoting by $Z_{e}=d^{N-m}$ 
the partition function of the $N-m$ environment degrees of freedom, we can rewrite the above as
\begin{eqnarray}
&&\sum_{m=m_{0},S_{z},l,e}^{N}w_{m}S_{z,e}^{2}\frac{e^{-\beta E_{l}^{m}}}{\tilde{Z}_{m}}\nonumber\\
&=& \sum_{m=m_{0},S_{z},l}^{N}w_{m} {\rm Tr}_{e}\left[\frac{S_{z,e}^{2}}{Z_{e}}\right]\frac{e^{-\beta E_{l}^{m}}}{{Z}_{m}}\nonumber\\
&=&\sum_{m=m_{0},l}^{N}w_{m}f_{2}(S)\frac{N-m}{8}B_{l}^{m}\nonumber
\end{eqnarray}
where $f_{2}(S)=\sum_{S_{z}}=(2S+1)$  and we used the fact that ${\rm Tr}_{e} S_{z,e}^{2}/Z_{e}=(N-m)/8$ since
for one environment state $\langle S_{z,e_{i}}^{2}\rangle_{e_{i}} \equiv
{\rm Tr}_{e_{i}} S_{z,e_{i}}^{2}/Z_{e_{i}}= (1/4 + 1/4)/4=1/8$ (for $d=4$). Hence, we have
\begin{eqnarray}
\frac{k_{\rm B}T\chi_{\rm E}}{(g\mu_{B})^{2}} &=&
\sum_{m=m_{0},l}^{N}f_{2}(S)\frac{N-m}{8}w_{m}B_{l}^{m}\label{fdm-chi-env},
\end{eqnarray}
and the total susceptibility $\chi(T)=\chi_{\rm S}(T)+\chi_{\rm E}(T)$ is given by
\begin{eqnarray}
\frac{k_{\rm B}T\chi(T)}{(g\mu_{B})^{2}} 
&\equiv & \frac{k_{\rm B}T\chi_{\rm S}(T)}{(g\mu_{B})^{2}} + \frac{k_{\rm B}T\chi_{\rm E}(T)}{(g\mu_{B})^{2}}\nonumber\\ 
&=&\sum_{m=m_{0},l}^{N}(f_{1}(S)+f_{2}(S)\frac{N-m}{8})w_{m}B_{l}^{m}\label{fdm-chi},
\end{eqnarray}
with a similar expression for the host susceptibility $\chi_{0}(T)$. The impurity contribution
is then obtained via $\chi_{\rm imp}(T)=\chi(T) -\chi_{0}(T)$. 

\begin{figure}
\includegraphics[width=\linewidth,clip]{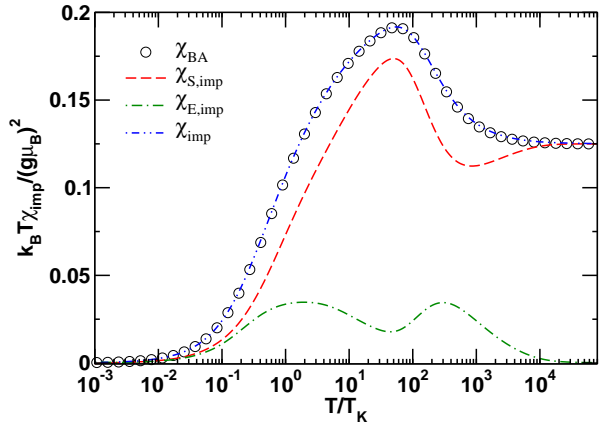}
\caption
{
  {\em (Color online)} 
  Impurity contribution to the susceptibility from Bethe ansatz 
  ($\chi_{\rm BA}$) and FDM ($\chi_{\rm imp}$) vs
  $T/T_{\rm K}$ for the symmetric Anderson model 
  [$U/\Delta_{0}=12$ and $\Delta_{0}=0.001D$ with $T_{\rm K}$ 
    defined in Eq.~(\ref{eq:symmetric-TK})]. Also shown are the
  impurity contributions $k_{\rm B}T\chi_{\rm S, imp}(T)/(g\mu_{\rm B})^2$ 
  and $k_{\rm B}T\chi_{\rm E,imp}(T)/(g\mu_{\rm B})^2$ as defined at the end of Sec.~\ref{sec:fdm}.
  The calculations are for $\Lambda = 10$ with an  energy 
  cut-off $e_{c}(\Lambda)=15\sqrt{\Lambda}\approx 47$, with $z$-averaging  
  [$n_{z}=4$, $z=1/8,\, 1/2,\, 3/8,\, 3/4$].
  \label{fdmfig1}
}
\end{figure}

Figure~\ref{fdmfig1} illustrates the problem just discussed for the symmetric Anderson
model in the strong correlation limit ($U/\Delta_{0}=12\gg 1$). 
Denoting by $\chi_{\rm S, imp}$ and $\chi_{\rm E, imp}$
the {\em impurity} contributions to $\chi_{\rm S}$ and $\chi_{\rm E}$, 
i.e. with respective host contributions subtracted, we have $\chi_{\rm imp}\equiv \chi_{\rm S, imp} 
+ \chi_{\rm E, imp}$. 
We see from Fig.~\ref{fdmfig1} that the contribution from the environment degrees of 
freedom, $\chi_{\rm E, imp}$, is significant at all temperatures and is required 
in order to recover the exact Bethe ansatz result for $\chi_{\rm imp}$. 

\section{Results}
\label{sec:results}
In this section we compare results for the impurity specific heat 
(Sec.~\ref{subsec:specific heat}), 
impurity susceptibilities in response to uniform (Sec.~\ref{subsec:susceptibility}) and
local  (Sec.~\ref{subsec:local susceptibility}) magnetic fields,
and the double occupancy (Sec.~\ref{subsec:double occupancy}) 
of the Anderson model, calculated within the FDM approach, 
with corresponding results from the conventional approach. 
For the first two quantities we also
show comparisons with Bethe ansatz calculations. 
Results for the Wilson ratio, as a function of Coulomb interaction
and local level position, within FDM and Bethe ansatz, are also presented 
(Sec.~\ref{subsec:susceptibility}). We show the results for all quantities
as functions of the reduced temperature $T/T_{\rm K}$, where the Kondo scale 
$T_{\rm K}$ is chosen to be the symmetric Anderson model Kondo scale given by
\begin{equation}
\label{eq:symmetric-TK}
T_{K}=\sqrt{U\Delta_{0}/2}e^{-\pi U/8\Delta_{0} + \pi \Delta_{0}/2U},
\end{equation}
except for $U/\Delta_{0}<1$ when we set $T_{K}=\Delta_{0}$. The Kondo scale in 
Eq.~(\ref{eq:symmetric-TK}) is related to
the $T=0$ Bethe ansatz susceptibility $\chi_{\rm imp}(0)$ via 
$\chi_{\rm imp}(0)=(g \mu_{\rm B})^{2}/4T_{\rm K}$ in the 
limit $U\gg\Delta_{0}$ (see Ref.~\onlinecite{Hewson1997}). We continue to
use $T_{\rm K}$ in comparing results from different methods, although we note that
for the asymmetric Anderson model, the physical low energy Kondo scale, $T_{\rm L}$, 
will increasingly deviate from $T_{\rm K}$ with increasing level asymmetry 
$\delta=2\varepsilon_{d}+U$. For example, second order poor Man's scaling 
for the Anderson model yields a low energy scale \cite{Haldane1978} 
$T_{\rm L}=\sqrt{U\Delta_{0}/2}e^{\pi\varepsilon_{d}(\varepsilon_{d}+U)/2\Delta_{0}U}$.
\subsection{Specific heat}
\label{subsec:specific heat}
\begin{figure}
\includegraphics[width=\linewidth,clip]{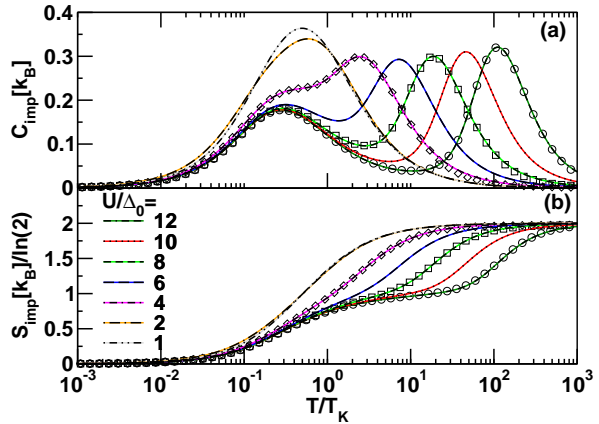}
\caption
{
  {\em (Color online)} (a) Impurity specific heat, $C_{\rm imp}(T)$, and, (b),  
  impurity entropy, $S_{\rm imp}(T)$, vs reduced temperature $T/T_{\rm K}$ 
  for the symmetric Anderson model with $U/\Delta_{0}=12,10,8,6,4,2,1$
  and $\Delta_{0}=0.001D$.
  Broken lines: FDM approach. Solid lines: conventional approach. 
  Selected Bethe ansatz results are shown as symbols for $U/\Delta_{0}=12,8,4$ (circles,
  squares, diamonds, respectively).
  The Kondo scale  is defined in Eq.~(\ref{eq:symmetric-TK}).
{As a guide to the eye, note that the high temperature 
  peak in $C_{\rm imp}$ shifts downwards with decreasing $U$.
}
  NRG and $z$-averaging parameters as in Fig.~\ref{fdmfig1}.
  \label{fdmfig2}
}
\end{figure}

\begin{figure}
\includegraphics[width=\linewidth,clip]{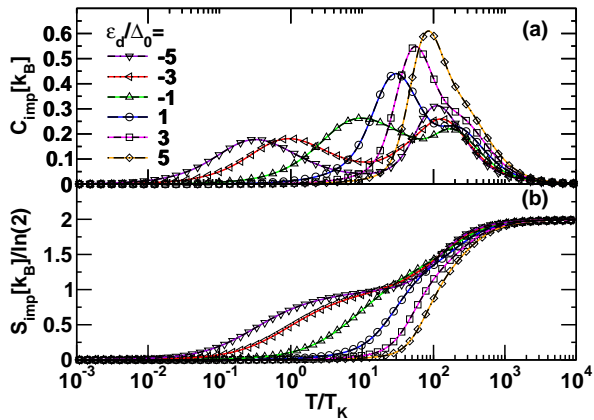}
\caption
{
  {\em (Color online)} 
  (a) Impurity specific heat, $C_{\rm imp}(T)$, and, (b),
  impurity entropy $S_{\rm imp}(T)$, vs reduced temperature $T/T_{\rm K}$ 
  for the asymmetric Anderson
  model with $U/\Delta_{0}=12$, $\Delta_{0}=0.001D$ and several values of 
  $\varepsilon_{d}/\Delta_{0}=-5,-3,\dots,+5$. 
  Broken lines: FDM approach. Solid lines: conventional approach. 
  Bethe ansatz results are shown as symbols for $\varepsilon_{d}/\Delta$. 
  For simplicity we used the symmetric $T_{\rm K}$ of Eq.~(\ref{eq:symmetric-TK}) for
  all $\varepsilon_{d}$ values. 
  NRG and $z$-averaging parameters as in Fig.~\ref{fdmfig1}.
  \label{fdmfig3}
}
\end{figure}

Figure~\ref{fdmfig2} shows the impurity specific heat ($C_{\rm imp}$) 
and impurity entropy ($S_{\rm imp}$) for the symmetric Anderson model
versus temperature $T/T_{\rm K}$ for increasing Coulomb interaction $U/\Delta_{0}$. 
One sees from Fig.~\ref{fdmfig2} that there
is excellent agreement between the results obtained within the FDM approach, within
the conventional approach and within the exact Bethe ansatz calculations. This agreement
is found for both the strongly correlated limit $U/\Delta_{0}\gg 1$ where there are 
two peaks in the specific heat, a low temperature Kondo induced peak and a high temperature
peak due to the resonant level, and for the weakly correlated limit $U/\Delta_{0} \lesssim 1$,
where there is only a single resonant level peak in the specific heat. The correct high 
temperature entropy $\ln 4$ is obtained in all cases. 

The temperature dependence of the impurity specific heat and entropy, for
the asymmetric Anderson model is shown in Fig.~\ref{fdmfig3} for local level positions
ranging from $\varepsilon_{d}/\Delta_{0}=-6$ to $+5$ in units of $\Delta_{0}$. For
simplicity we continue to show the results as a function of $T/T_{\rm K}$, with $T_{\rm K}$
the symmetric Kondo scale (\ref{eq:symmetric-TK}), although, the true Kondo scale will
deviate from this for $\varepsilon_{d}> -U/2$. The FDM results agree also here 
very well with the conventional approach and the Bethe ansatz calculations. 

\subsection{Susceptibility and Wilson ratio}
\label{subsec:susceptibility}
\begin{figure}
\includegraphics[width=\linewidth,clip]{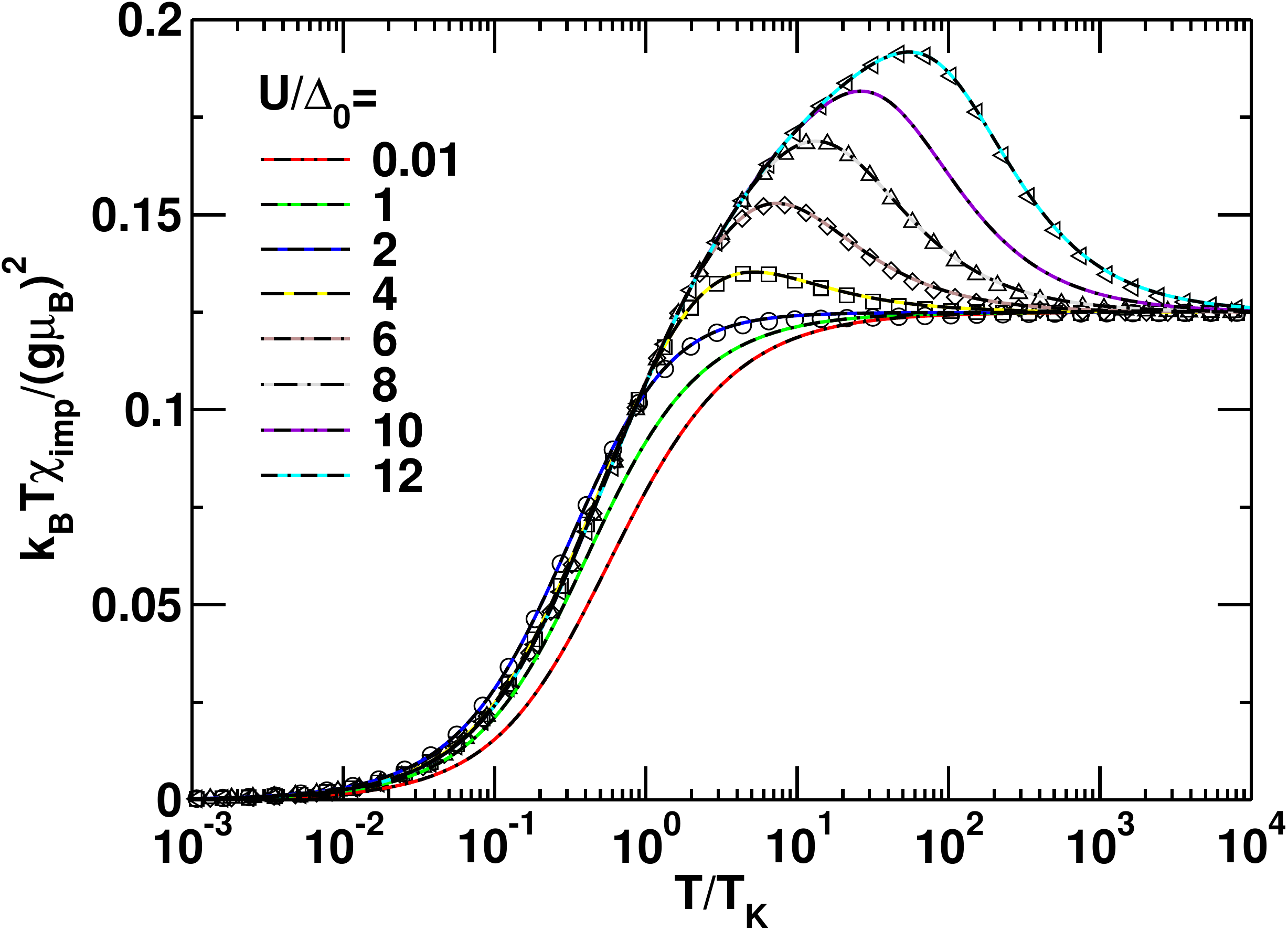}
\caption
{
  {\em (Color online)} 
  Impurity susceptibility, $\chi_{\rm imp}(T)$, vs $T/T_{\rm K}$ 
  for the symmetric Anderson model with $U/\Delta_{0}=12,10,8,6,4,2,1,0.01$ 
  and $\Delta_{0}=0.001D$ with $T_{\rm K}$ defined in 
  Eq.~(\ref{eq:symmetric-TK}) for $U/\Delta_{0}\ge 1$ and $T_{K}=\Delta_{0}$ for the 
  $U/\Delta_{0}=0.01$ case. 
  Broken lines: FDM approach. Solid lines: conventional approach. Symbols: Bethe ansatz
  (for selected values of $U/\Delta_{0}=12,8,6,4,2$). 
{As a guide to the eye, note that
  $\chi_{\rm imp}$ is increasingly enhanced with increasing $U$.
} 
  NRG and $z$-averaging parameters as in Fig.~\ref{fdmfig1}.
  \label{fdmfig4}
}
\end{figure}

Fig.~\ref{fdmfig4} compares the susceptibilities of the symmetric 
Anderson model calculated from FDM, conventional and
Bethe ansatz approaches for several values of $U/\Delta_{0}$, again
indicating good agreement over the whole temperature range between these
three approaches. 
\begin{table}
\begin{ruledtabular}
\begin{tabular}{d|dd|dd}
\multicolumn{1}{c|}{\multirow{2}{*}{$U/\Delta_0$}} &  \multicolumn{2}{c|}{$k_B T_K \chi_\mathrm{imp}^a/(g \mu_B)^2$} &\multicolumn{2}{c}{$R^a$}\\
&\multicolumn{1}{c}{$a = \mathrm{BA}$}&\multicolumn{1}{c|}{$a = \mathrm{NRG}$}&\multicolumn{1}{c}{$a = \mathrm{BA}$}&\multicolumn{1}{c}{$a = \mathrm{NRG}$} \\ \tableline
12 & 0.250091 & 0.256 &  1.998 &2.027\\
10 & \text{--} &  0.256 & \text{--} &2.024\\
8 & 0.250715 & 0.256 & 1.986 & 2.013\\
6 & \text{--} &  0.2574 &\text{--} & 1.982\\
4 &  0.259130 &  0.2637 & 1.852 &1.877\\
2 & \text{--} & 0.3085 & \text{--} & 1.578\\
1 & \text{--} &  0.2214 & \text{--} &1.317\\
0.01 & \text{--} & 0.1599 & \text{--} & 1.003
\end{tabular}
\end{ruledtabular}
\caption{Zero temperature susceptibilities 
$k_{\rm B}T_{K}\chi_{\rm imp}^{a}/(g\mu_{\rm B})^{2}$ and Wilson ratios $R^{a}\equiv\lim_{T\rightarrow 0}4\pi^{2}\chi_{\rm imp}^{a}(T)/3 C_{\rm imp}^{a}(T)/T$ for the symmetric Anderson model
at several values of $U/\Delta_{0}$ using the Bethe ansatz/NRG FDM approach ($a={\rm BA/NRG}$). 
Note that $T_{\rm K}$ is defined by 
Eq.~\ref{eq:symmetric-TK} for $U/\Delta_{0}> 1$ and is set to $\Delta_{0}$ otherwise.
}
\label{TableI}
\end{table}
Table~\ref{TableI} lists the zero temperature impurity susceptibilities ($k_{\rm B}T_{K}\chi_{\rm imp}/(g\mu_{\rm B})^{2}$)
and Wilson ratios ($R\equiv\lim_{T\rightarrow 0}4\pi^{2}\chi_{\rm imp}(T)/3 C_{\rm imp}(T)/T$),
for the symmetric Anderson model as calculated within FDM and for 
a range of Coulomb interactions from strong $U/\Delta_{0}\gg 1$ to 
weak ($U/\Delta_{0}\ll 1$). In these two limits, the Wilson ratio 
for the symmetric Anderson model approaches the
well known values of $2$, and $1$, respectively within FDM ($a=NRG$) and Bethe ansatz. Comparison with Bethe ansatz 
results at selected values of $U/\Delta_{0}$ indicate an error in the 
susceptibility of around $2\%$ with a similar error in the Wilson ratio.

\begin{figure}
\includegraphics[width=\linewidth,clip]{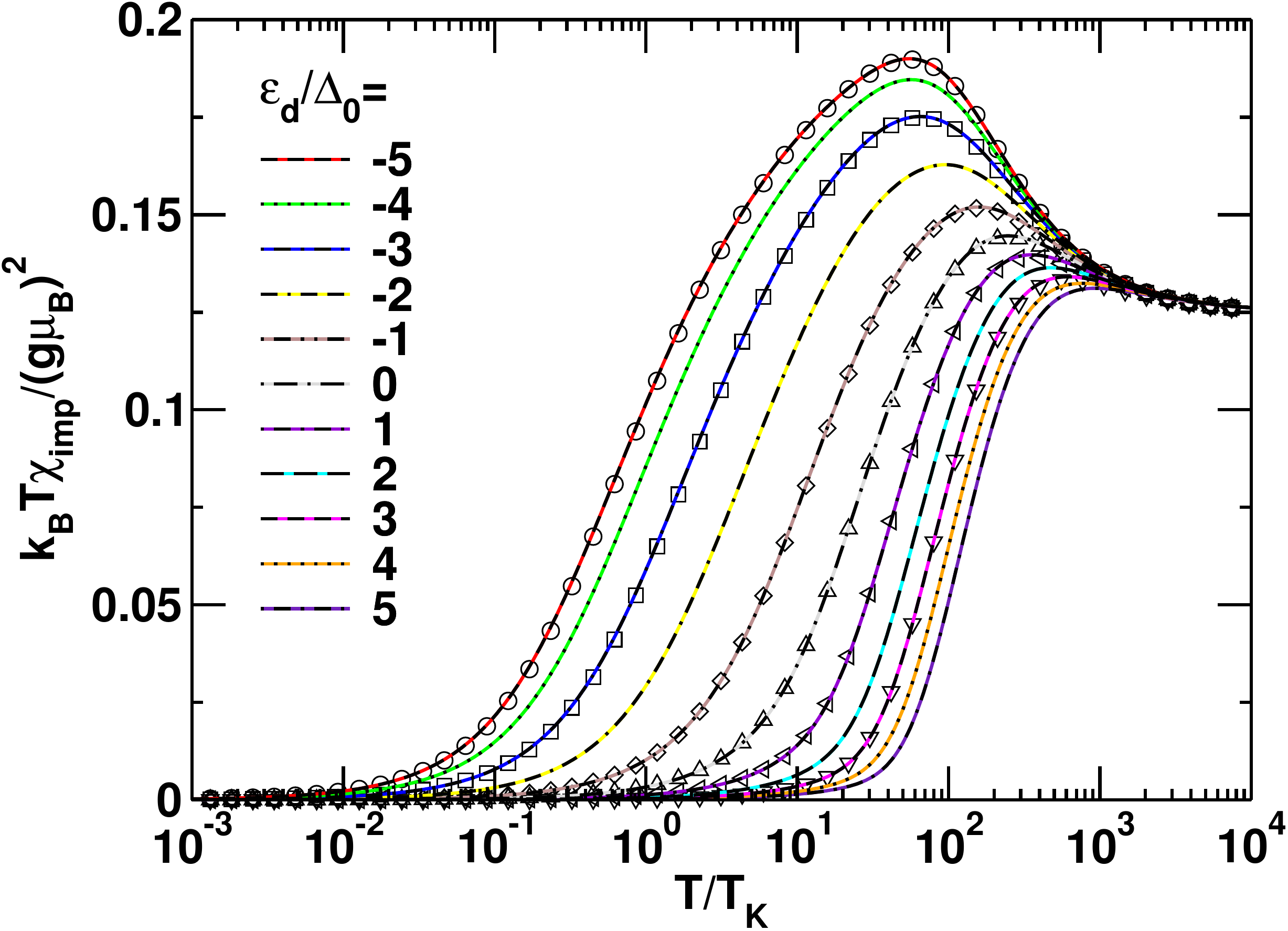}
\caption
{
  {\em (Color online)} 
  Impurity susceptibility, $\chi_{\rm imp}(T)$, vs $T/T_{\rm K}$ for the asymmetric Anderson
  model with $U/\Delta_{0}=12$, $\Delta_{0}=0.001D$ and several values of $\varepsilon_{d}/\Delta_{0}$
  with $T_{\rm K}$ defined in Eq.~(\ref{eq:symmetric-TK}). 
  Broken lines: FDM approach. Solid lines: conventional approach. Symbols: Bethe ansatz (for
  selected values of $\varepsilon_{d}/\Delta_{0}=-5,-3,-1,0,+1,+3$). 
{As a guide to the eye,
  note that the susceptibility curves shift to higher temperatures with increasing $\varepsilon_{d}$.
}
  NRG and $z$-averaging parameters as in Fig.~\ref{fdmfig1}.
  \label{fdmfig5}
}
\end{figure}
Fig.~\ref{fdmfig5} shows results within FDM and conventional approaches
for the asymmetric Anderson model ($\varepsilon_{d}> -U/2$) and
for several local level positions ranging from the Kondo 
($-\varepsilon_{d}/\Delta_{0}\gg 1$) to the mixed valence 
$|\varepsilon_{d}/\Delta_{0}|\le 1$ and empty orbital 
regimes $\varepsilon_{d}/\Delta_{0}> 1$. Bethe ansatz results are
also shown for selected local level positions, and we see again very 
good agreement between all three methods over the whole temperature 
range. Corresponding zero temperature susceptibilities and Wilson
ratios are listed in Table~\ref{TableII}. Note that the Wilson ratio
approaches the value for a non-interacting system only in the empty orbital
limit ($\varepsilon_{d}\gg \Delta_{0}$), being approximately $1.5\pm 0.25$ in
the mixed valence regime ($|\varepsilon_{d}/\Delta_{0}|\lesssim 1$). 
The Wilson ratio from NRG and Bethe ansatz deviate by less than $3\%$ 
in all regimes. 
\begin{table}
\begin{ruledtabular}
\begin{tabular}{d|dd|dd}
\multicolumn{1}{c|}{\multirow{2}{*}{$\varepsilon_{d}/\Delta_0$}} &  \multicolumn{2}{c|}{$k_B T_K \chi_\mathrm{imp}^a/(g \mu_B)^2$} &\multicolumn{2}{c}{$R^a$} \\ 
& \multicolumn{1}{c}{$a=BA$}&\multicolumn{1}{c|}{$a=NRG$}& \multicolumn{1}{c}{$a=BA$}&\multicolumn{1}{c}{$a=NRG$} \\ \tableline
-5 & 0.219482 &  0.2245 & 1.999 & 2.025\\
-4 & \text{--}&  0.1515 &\text{--} & 2.023\\
-3 & 0.077356 &  0.0785 & 1.990 &2.00\\
-2 & \text{--} &  0.0315 & \text{--} &1.97\\
-1 & 0.010337 &   0.0103 & 1.795& 1.78\\
0 & 0.003303 &0.0033 &  1.512 & 1.50\\
1 & 0.001250 &  0.0013 &1.315 & 1.32\\
2 & \text{--} &  0.00059 & \text{--} &1.18\\
3 & 0.000325 &  0.00033 &1.086 & 1.12\\
4 & \text{--} & 0.00021 & \text{--} & 1.09\\
5 & \text{--} &  0.00014 & \text{--} &1.06
\end{tabular}
\end{ruledtabular}
\caption{Zero temperature susceptibilities 
$k_{\rm B}T\chi_{\rm imp}^{a}/(g\mu_{\rm B})^{2}$ and Wilson ratios $R^{a}\equiv\lim_{T\rightarrow 0}4\pi^{2}\chi_{\rm imp}^{a}(T)/3 C_{\rm imp}^{a}(T)/T$ for the asymmetric Anderson model at $U/\Delta_{0}=12$ and several local level positions $\varepsilon_{d}/\Delta_{0}$ 
using the Bethe ansatz/FDM NRG approach ($a={\rm BA/NRG}$). 
}
\label{TableII}
\end{table}

%
%
\subsection{Local susceptibility}
\label{subsec:local susceptibility}
It is also interesting to consider the susceptibility, $\chi_{\rm loc}$, 
in response to a local magnetic field acting only at the impurity site and to compare
this with the susceptibility, $\chi_{\rm imp}$, discussed above, in which the magnetic 
field acts on both the impurity and conduction electron spins. The former is relevant, for
example, in nuclear magnetic resonance and neutron scattering experiments, while the
latter can be measured in bulk samples with and without magnetic impurities.

{A local magnetic field term, $-g\mu_{\rm B}BS_{\rm z,d}$, in the Anderson model, with
$S_{z,d}=(n_{d\uparrow}-n_{d\downarrow})/2$, is not a conserved
quantity, i.e  $S_{\rm z,d}$ is not conserved
}
, and $\chi_{\rm loc}(T)$ cannot be expressed
as a fluctuation as in Eq.~(\ref{chi-total}-\ref{chi-host}),  which would obviate the
need to explicitly evaluate a numerical second derivative with respect to $B$ of the
thermodynamic potential. Such a derivative, however, poses no actual problem within
NRG, so we proceed by explicitly diagonalizing the Anderson model in a local field, using 
only ${\rm U}(1)$ symmetries for charge and spin (for the symmetric Anderson model in a 
magnetic field, an ${\rm SU}(2)$ pseudo-spin symmetry may be exploited, by using the mapping of this 
model in a local magnetic field $B$ onto the ${\rm SU}(2)$ invariant negative-$U$ 
Anderson model in zero magnetic field at finite level asymmetry $2\varepsilon_{d}+U=B$
\cite{Iche1972,Hewson2006}). The evaluation of $\chi_{\rm loc}$ then proceeds via  
$\chi_{\rm loc}(T,B=0)= -
\partial^{2}\Omega_{\rm loc}(T,B)/\partial B^{2}|_{B=0}$ where 
$\Omega_{\rm loc}(T,B) = \Omega(T,B)-\Omega_{\rm 0}(T)$ and
$\Omega(T,B)$ and $\Omega_{\rm 0}(T)$ are the thermodynamic potentials of 
the total system in a local magnetic field $B$ and the host system, respectively.

\begin{figure}
\includegraphics[width=\linewidth,clip]{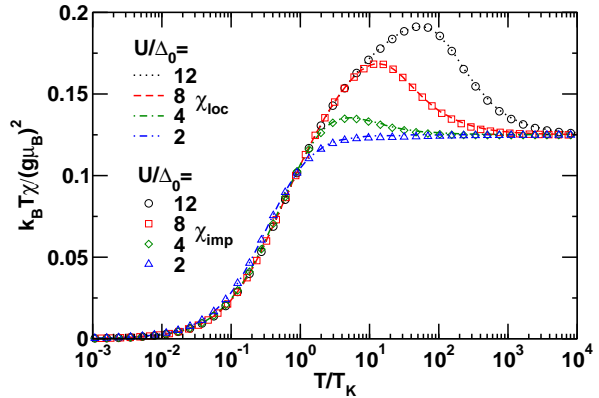}
\caption
{
  {\em (Color online)} 
  Comparison of the local, $\chi_{\rm loc}(T)$, and impurity, $\chi_{\rm imp}$, 
  susceptibilities vs $T/T_{\rm K}$ 
  for the symmetric Anderson model with $U/\Delta_{0}=12,8,4,2$ 
  and $\Delta_{0}=0.001D$ with $T_{\rm K}$ defined in 
  Eq.~(\ref{eq:symmetric-TK}).
  Broken lines: FDM approach. Symbols: Bethe ansatz
  (for selected values of $U/\Delta_{0}=12,8,4,2$).
  NRG and $z$-averaging parameters as in Fig.~\ref{fdmfig1}.
  \label{fdmfig6}
}
\end{figure}

Results for $\chi_{\rm loc}$ obtained in this way are shown in Fig.~\ref{fdmfig6} at several
values of $U/\Delta_{0}$ as a function of $T/T_{\rm K}$. A comparison of $\chi_{\rm loc}$ to
$\chi_{\rm imp}$ obtained from the Bethe ansatz, allows us to conclude that these two 
susceptibilities are close to identical at all temperatures, i.e. $\chi_{\rm imp}(T)=\chi_{\rm loc}(T)$
and for all interaction strengths $U/\Delta_{\rm 0}$. This is not always the case. A
prominent example is the anisotropic Kondo model \cite{Vigman1978}, where 
$\chi_{\rm imp}=\alpha \chi_{\rm loc}$, with the dissipation strength $0\leq \alpha\leq 1$ 
being determined by the anisotropy of the exchange interaction \cite{Vigman1978,Hakim1985}.

\begin{figure}
\includegraphics[width=\linewidth,clip]{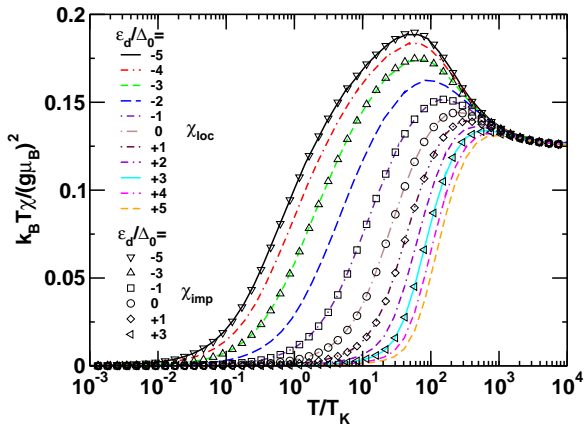}
\caption
{
  {\em (Color online)} 
  Comparison of the local, $\chi_{\rm loc}$, and impurity, $\chi_{\rm imp}(T)$,
  susceptibilities vs $T/T_{\rm K}$ for the asymmetric Anderson model with $U/\Delta_{0}=12$, 
  $\Delta_{0}=0.001D$ with $T_{\rm K}$ defined in 
  Eq.~(\ref{eq:symmetric-TK}) and for several values of the local level position:
  $\varepsilon_{d}/\Delta_{\rm 0}=-5,-4,\dots,+5$. 
  Broken lines: $\chi_{\rm loc}$ from the FDM approach. Symbols: $\chi_{\rm imp}$ from 
  Bethe ansatz for selected values of $\varepsilon_{d}$.
  The NRG calculations are for $\Lambda=4$ with an energy cut-off $e_{c}(\Lambda=4)=40$ with
  $z$-averaging [$n_{z}=2$ with $z=0$ and $z=0.5$].
  \label{fdmfig7}
}
\end{figure}

Figure~\ref{fdmfig7} compares local and impurity susceptibilities for the
asymmetric Anderson model in the strong correlation limit ($U/\Delta_{\rm 0}=12$)
for several local level positions, ranging from the Kondo ($\varepsilon_{d}/\Delta_{0}=-5,-4,-3,-2$) 
to the mixed valence ($\varepsilon_{d}/\Delta_{0}=-1,0,+1$) 
and into the empty orbital regime ($\varepsilon_{d}/\Delta_{0}=+2,\dots,+5$).  
We see that, as for the symmetric Anderson model, 
local and impurity susceptibilities are almost identical at all temperatures and for
all local level positions, i.e. $\chi_{\rm imp}(T)=\chi_{\rm loc}(T)$ for the parameter values
used. 

The result $\chi_{\rm imp}(T)=\chi_{\rm loc}(T)$, which we verified here, follows from the
Clogston-Anderson compensation theorem\cite{Clogston1961} (see Ref.~\onlinecite{Hewson1997}). Consider the
impurity contribution to the magnetization $M_{\rm imp}(B)$ in a uniform field. This is 
given by $M_{\rm imp}/(g\mu_{\rm B})=\langle S_{z,d} + S_{z,c}\rangle -\langle S_{z,c}\rangle_{\rm 0}$ 
where $S_{z,d}, S_{z,c}$ are the impurity and conduction electron $z-$components of spin. Using
equations of motion, one easily shows \cite{Hewson1997}, that the additional impurity magnetization
$\delta M_{\rm imp}/(g\mu_{\rm B})=\langle S_{z,c}\rangle -\langle S_{z,c}\rangle_{\rm 0}$ from the
conduction electrons induced by the presence of the impurity is given by 
\begin{eqnarray}
&&\frac{\delta M_{\rm imp}}{(g\mu_{\rm B})} 
= \frac{1}{2\pi}\sum_{\sigma}\int d\omega f(\omega) 
{\rm Im}\left[\sigma G_{d\sigma}(\omega,B)
\frac{\partial \Delta(\omega)}{\partial\omega}\right],\label{eq:magnetization-correction}
\end{eqnarray}
where $f(\omega)$ is the Fermi function, $G_{d\sigma}(\omega,B)$ is the spin $\sigma$ local
level Green function of the Anderson model and 
$\Delta(\omega)=\sum_{k}|V_{k}|^{2}/(\omega-\epsilon_{k}+i\delta)$ is the hybridization function. For a 
flat band, $\partial \Delta(\omega)/\partial\omega \approx \Delta_{0}/D$ in the wide-band limit. 
Hence, $\delta M_{\rm imp}/(g\mu_{\rm B})$ is of order $(M_{\rm loc}(B)/g\mu_{\rm B})\Delta_{0}/D$
where $M_{\rm loc}(B)/(g\mu_{\rm B})=\langle n_{d\uparrow}-n_{d\downarrow}\rangle/2 = \langle S_{z,d}\rangle$ is the local 
magnetization, which is linear in $B$ for $B\rightarrow 0$. From this we deduce that 
 $M_{\rm imp}/(g\mu_{\rm B})\approx \langle S_{z,d}\rangle\equiv M_{\rm loc}/(g\mu_{\rm B})$ to within
corrections of order $(M_{\rm loc}(B)/g\mu_{\rm B})\Delta_{0}/D$, i.e. 
$\chi_{\rm imp}(T)=\chi_{\rm loc}(T)$ to within corrections of order 
$(\chi_{\rm loc}(T)/(g\mu_{\rm B})^{2})\Delta_{0}/D\ll \chi_{\rm loc}(T)$. Away from the wide-band limit,
or for strong energy dependence of $\Delta(\omega)$, the above susceptibilities will differ by the
correction term given by the field derivative of $\delta M_{\rm imp}$ in 
Eq.~(\ref{eq:magnetization-correction}).

\subsection{Double occupancy}
\label{subsec:double occupancy}
\begin{figure}
\includegraphics[width=\linewidth,clip]{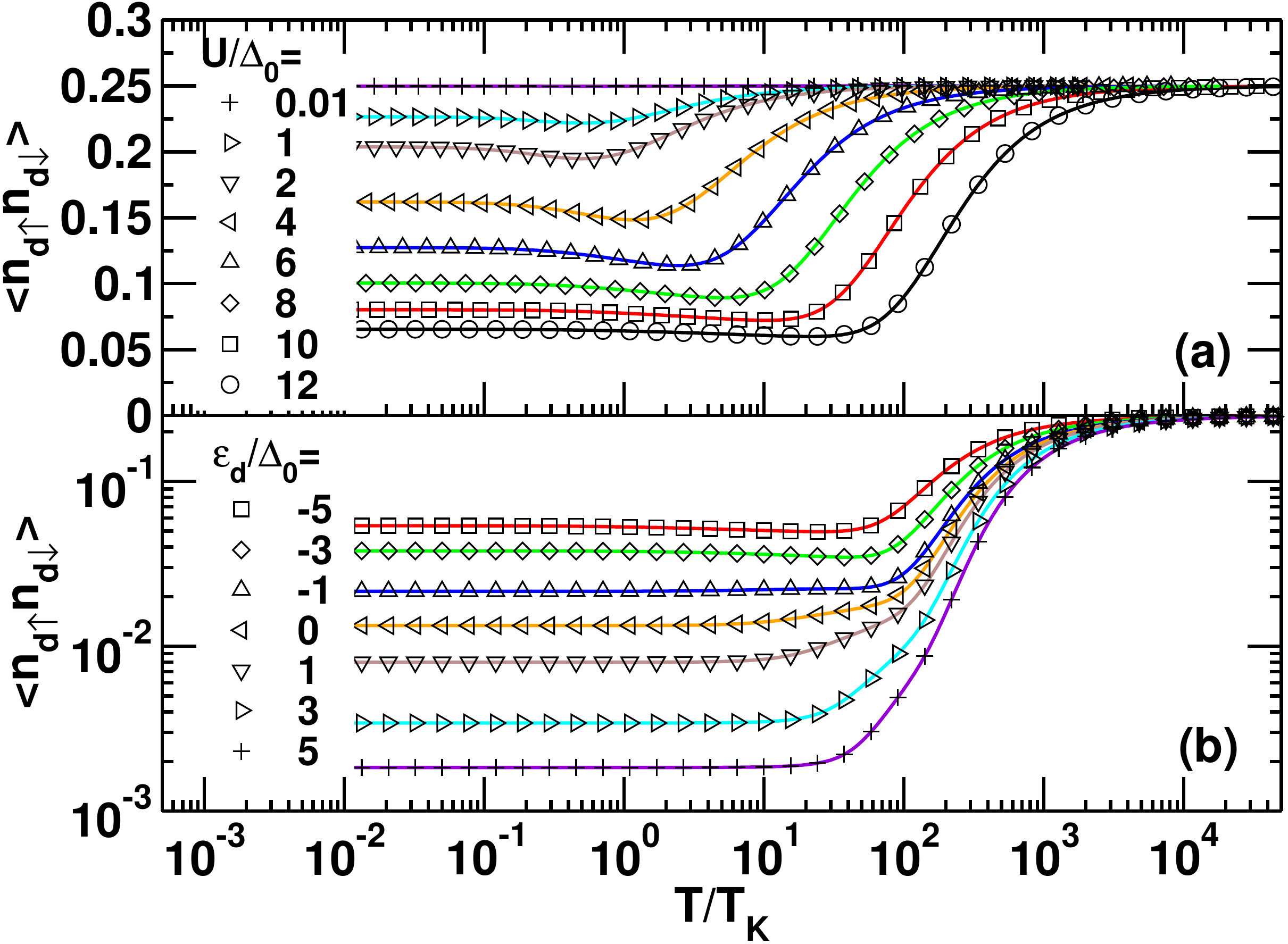}
\caption
{
  {\em (Color online)} 
  (a) Double occupancy $D_{\rm occ}=\langle n_{d\uparrow}n_{d\downarrow}\rangle $ as a function of temperature
      $T/T_{\rm K}$  for the symmetric Anderson model and decreasing values of 
  $U/\Delta_{0}=12,10,8,6,4,2,1,0.01$ within FDM (solid lines) and conventional approaches (symbols).
   (b) Double occupancy $D_{\rm occ}=\langle n_{d\uparrow}n_{d\downarrow}\rangle $ as a function of temperature
  $T/T_{\rm K}$  for the asymmetric Anderson model and increasing values of 
  $\varepsilon_{d}/\Delta_{0}=-5,-3,-1,0,+1,+3,+5$ for $U/\Delta_{0}=12$ 
  within FDM (solid lines) and conventional approaches (symbols).
  $T_{\rm K}$ is defined in Eq.~(\ref{eq:symmetric-TK}) for $U/\Delta_{0}\ge 1$ and is set to $\Delta_{0}$
  for the case $U/\Delta_{0}=0.01$. 
  NRG and $z$-averaging parameters as in Fig.~\ref{fdmfig1}.
  \label{fdmfig8}
}
\end{figure}

Our conclusions concerning the accuracy of specific heat and the susceptibility
calculations within the FDM approach, hold also for 
other thermodynamic properties, e.g. for the occupation number or the double occupancy. 
Fig.~\ref{fdmfig8}(a) shows a comparison between
the FDM and conventional approaches for the temperature dependence of the double occupancy 
{
$D_{\rm occ}=\langle n_{d\uparrow}n_{d\downarrow}\rangle $
} 
of the symmetric model for different strengths
of correlation $U/\Delta_{0}$, and  Fig.~\ref{fdmfig8}(b) shows the same for the asymmetric
Anderson model for $U/\Delta_{0}=12$ and for several local level positions. The results of the
two approaches agree at all temperatures, local level positions and Coulomb interactions. 
Notice that $D$ acquires its mean-field value of $1/4$ for the symmetric model in the
limit $U/\Delta_{0}\rightarrow 0$ and is strongly suppressed with increasing Coulomb 
interaction away from this limit [see Fig.~\ref{fdmfig8}(a)]. Similarly for the asymmetric
model, increasing $\varepsilon_{d}/\Delta_{0}$ away from the correlated Kondo regime decreases
the double occupancy significantly [see Fig.~\ref{fdmfig8}(b)].

\section{Summary}
\label{sec:summary}
In this paper we focused on the calculation of the impurity specific heat and 
the impurity susceptibility
of the Anderson model within the FDM approach \cite{Weichselbaum2007}, 
finding that this method gives reliable results for
these quantities, as shown by a comparison to both exact Bethe ansatz calculations 
\cite{Kawakami1981,Kawakami1982a,Okiji1983,Wiegmann1983,Tsvelick1983,Filyov1982,Tsvelick1982}
and to NRG calculations within the conventional approach \cite{Campo2005}. Some care is needed
in implementing the FDM approach for the susceptibility $\chi_{\rm imp}$ in a uniform field, i.e. when 
the applied magnetic field acts on both the impurity and conduction electron spins. 
In this case, an additional contribution from the environment degrees of freedom needs 
to be included. We also showed that the susceptibility in response to a local magnetic field
on the impurity, $\chi_{\rm loc}$, could also be obtained within FDM and a comparison of this
susceptibility with $\chi_{\rm imp}$ (from Bethe ansatz), showed that they are close to identical at all 
temperatures, and in all parameter regimes for $\Delta_{0}\ll D$, thereby verifying the Clogston-Anderson
compensation theorem.  
An arbitrary temperature grid can be used for thermodynamics in both the 
conventional and the FDM approaches, however, the former requires a specific best shell 
to be selected depending on $T$ and $z$, whereas the FDM approach avoids this step by
incorporating all excitations from all shells in a single density matrix. 

We also showed, that quantities such as the double occupancy can also be
accurately calculated within the FDM approach. The double occupancy can be 
probed in experiments on cold atom realizations of Hubbard models in optical 
lattices. \cite{Joerdens2008,Schneider2008} Flexible techniques,
such as the FDM approach, for calculating them within a dynamical mean field theory 
\cite{Kotliar2004,Georges1996,Vollhardt2012} treatment of the underlying 
effective quantum impurity models could be useful in future investigations 
of such systems \cite{Tang2012}.

\begin{acknowledgments}
We thank Jan von Delft, Markus Hanl and Ralf Bulla for useful discussions and 
acknowledge supercomputer support by the John von Neumann institute for Computing (J\"ulich).
Support from the DFG under grant number WE4819/1-1 is also acknowledged (AW).
\end{acknowledgments}
\appendix
\bibliography{spec}
\end{document}